
\input tables.tex
\input phyzzx.tex
\def\yhmach{y_{\h}^{mach}}
\def\xmin{x_{min}}
\def\naiii{N_{\cala_3}}

\def\nsdaiii{N_{SD}({\cala_3})}
\def\yhopt{y_{\h}^{opt}}
\def\eepem{E_{\epem}}
\def\eepemopt{E_{\epem}^{opt}}
\def\eepemmach{E_{\epem}^{mach}}

\def\hpm{H^{\pm}}

\def\vev#1{\langle #1 \rangle}

\def\h{h}
\def\mh{m_{\h}}
\def\tanb{\tan\beta}

\def\hl{h^0}
\def\mhl{m_{\hl}}
\def\hh{H^0}
\def\mhh{m_{\hh}}
\def\hpm{H^{\pm}}

\def\mt{m_t}

\def\hl{h^0}
\def\hh{H^0}
\def\ha{A^0}
\def\mhl{m_{\hl}}
\def\mhh{m_{\hh}}
\def\mha{m_{\ha}}

\def\lam{\lambda}

\def\gam{\gamma}
\def\gam{\gamma}

\def\gam{\gamma}

\def\calm{{\cal M}}
\def\cala{{\cal A}}
\def\call{{\cal L}}
\def\stop{{\wtilde t}}
\def\mstop{m_{\stop}}
\def\ep{e^+}
\def\em{e^-}

\def\gev{~{\rm GeV}}
\def\tev{~{\rm TeV}}

\def\fbi{~{\rm fb}^{-1}}

\def\wt{\widetilde}


\def\prdj#1{{\it Phys. Rev.} {\bf D{#1}}}

\def\prlj#1{{\it Phys. Rev. Lett.} {\bf {#1}}}
\def\plbj#1{{\it Phys. Lett.} {\bf B{#1}}}

\def\mt{m_t}

\def\rta{\rightarrow}
\def\tanb{\tan\beta}

\def\cala{{\cal A}}

\def\calm{{\cal M}}

\def\nsd{N_{SD}}

\def\mz{m_Z}
\def\anti{\overline}

\def\ifmath#1{\relax\ifmmode #1\else $#1$\fi}

\def\3quarter{{\textstyle{3 \over 4}}}

\def\nou{\noindent\undertext}

\input phyzzx
\Pubnum={$\caps UCD-94-20$\cr}
\date{April, 1994}

\titlepage
\vskip 0.75in
\baselineskip 0pt
\hsize=6.5in
\vsize=8.5in
\centerline{{\bf Determining the CP-eigenvalues of the Neutral Higgs Bosons}}
\centerline{{\bf of the Minimal Supersymmetric Model in $\gam\gam$ Collisions}}
\vskip .075in
\centerline{ J.F. Gunion and J.G. Kelly}
\vskip .075in
\centerline{\it Davis Institute for High Energy Physics,
Dept. of Physics, U.C. Davis, Davis, CA 95616}

\vskip .075in
\centerline{\bf Abstract}
\vskip .075in
\centerline{\Tenpoint\baselineskip=12pt
\vbox{\hsize=12.4cm
\noindent We determine the optimal laser and $\ep/\em$ energies and
polarizations for {\it directly} determining the CP
eigenvalue of each of the neutral Higgs bosons of
the Minimal Supersymmetric Model via measurement of
transverse polarization cross section asymmetry in back-scattered
laser photon collisions.
Approximate statistical significances are computed for the measurement
of the CP eigenvalue as a function of Higgs mass and other
parameters of the model.
}}

\vskip .15in
\noindent{\bf 1. Introduction}
\vskip .075in

Supersymmetric models are of considerable interest as a
possibility for extending the Standard Model (SM).
\Ref\hhg{For a review see
J.F. Gunion, H.E. Haber, G.L. Kane, and S. Dawson, {\it The Higgs Hunter's
Guide}, Addison-Wesley, Redwood City, CA (1990).}\
A supersymmetric extension of the SM must contain at least two Higgs doublets.
The minimal supersymmetric model (MSSM) is defined by having precisely two
Higgs
doublets. The resulting physical Higgs sector contains two charged Higgs
bosons ($\hpm$), two CP-even Higgs bosons ($\hl,\hh$ with $\mhl<\mhh$), and one
CP-odd Higgs boson ($\ha$). (Note that CP is automatically conserved in the
MSSM
Higgs sector, by virtue of the SUSY constraints.) At tree-level, the MSSM Higgs
sector is entirely determined by just two parameters, which are normally
chosen to be $\tanb\equiv v_2/v_1$ (the ratio of the
vacuum expectation values of the neutral members of the Higgs doublets
that couple to up- and down-type quarks, respectively) and $\mha$.
One-loop radiative corrections are, however, important.
\Ref\radcorra{H. Haber and R. Hempfling \prlj{66} (1991) 1815; Y. Okada, M.
Yamaguchi and T. Yanagida, {\it Prog. Theor. Phys.} {\bf 85} (1991) 1;
J. Ellis, G. Ridolfi and F. Zwirner, \plbj{257} (1991) 83.}\
They depend most crucially upon the
top quark mass ($\mt$) and the stop squark mass ($\mstop$). However, other
parameters enter as well. These include the soft supersymmetry
breaking parameters,
$A_t,A_b$, and the $\mu$ parameter characterizing Higgs superfield mixing.

\REF\habgun{J.F. Gunion and H.E. Haber, Proceedings of the 1990 DPF
Summer Study on High Energy Physics, ``Research Directions
for the Decade'', Snowmass (1990), ed. E. Berger, p. 469; \prdj{48}
(1993) 2907.}
\REF\caldwelletal{D. Borden, D. Bauer, and D. Caldwell, \prdj{48} (1993)
4018. See also D. Borden, Proceedings of the ``Workshop on Physics
and Experiments with Linear $\epem$ Colliders'', Waikaloa, Hawaii,
April 26-30 (1993), eds. F.A. Harris, S.L. Olsen, S. Pakvasa, and X. Tata,
p. 323.}
Recently, attention has been given to the
possibility of forming Higgs bosons via
polarized photon collisions.\refmark{\habgun,\caldwelletal}\ Intense beams of
polarized photons can be produced by
back-scattering polarized laser beams off of
polarized electron and positron beams at a TeV-scale linear $\ep\em$ collider.
\REF\telnovi{H.F. Ginzburg, G.L. Kotkin, V.G. Serbo and V.I. Telnov,
{\it Nucl. Inst. and Meth.} {\bf 205}, 47 (1983).}
\REF\telnovii{H.F. Ginzburg, G.L. Kotkin, S.L. Panfil,
V.G. Serbo and V.I. Telnov,
{\it Nucl. Inst. and Meth.} {\bf 219}, 5 (1984).}
\refmark{\telnovi,\telnovii}\ In previous work,\refmark{\habgun}\
it has been established that
the neutral MSSM Higgs bosons can indeed
be detected in $\gam\gam$ collisions over
much of parameter space. In fact, since each can be produced singly by direct
$\gam\gam$ collisions, whereas they are only detectable in $\epem$ collisions
in the pair production mode,  $\epem\rta\ha\hh$, photon-photon colliders can
even have a larger mass reach in the case of the heavier $\ha$ and $\hh$ than
direct $\epem$ collisions.\refmark\habgun\
Once Higgs bosons are observed in either
$\epem$ collisions or $\gam\gam$ collisions, it will be crucial to check their
detailed properties. Since one of the most basic
features of the MSSM Higgs sector is the prediction that {\it pure} eigenstates
with CP$=+$ and CP=$-$ must both be present, it will be very important to make
a
direct determination of the CP of each Higgs boson that is detected.
Direct measurement of the CP-properties of the three neutral Higgs bosons
will, however, be challenging at any future collider.

A $\gam\gam$-collider allows a study of the
CP-properties of the Higgs sector by virtue of the
different couplings for two photons to CP-even vs. CP-odd Higgs bosons:
$$\calm[\gam\gam\rta \h(CP=++)]\propto {\vec \epsilon}\cdot {\wt{ \vec
\epsilon}}\,,\qquad \calm[\gam\gam\rta \h(CP=+-)]\propto
({\vec \epsilon}\times {\wt {\vec\epsilon}})_z\,,\eqn\couplings$$
where $\vec\epsilon$ and $\wt{\vec\epsilon}$ are the polarizations
of the two colliding photons in the $\gam\gam$ CM.
\REF\gg{J.F. Gunion and B. Grzadkowski, \plbj{294} (1992) 361.}
This was first exploited in Ref.~[\gg] to show that a Higgs boson of mixed CP
character could be distinguished
from a pure CP eigenstate by the fact that a certain asymmetry between
production cross sections initiated by colliding photons of different
helicities was non-zero.
\REF\gerice{J.F. Gunion, ``Properties of SUSY Particles'', Proceedings
of the 23rd Workshop of the INFN Eloisatron Project, Erice, Italy,
September 28 - October 4 (1992), eds. L. Cifarelli and V.A. Khoze,
p. 279.}
In contrast, it was noted in Refs.~[\gg] and [\gerice]
\REF\ghawaii{J.F. Gunion, Proceedings of the ``Workshop on Physics
and Experiments with Linear $\epem$ Colliders'', Waikaloa, Hawaii,
April 26-30 (1993), eds. F.A. Harris, S.L. Olsen, S. Pakvasa, and X. Tata,
p. 166.}
\REF\djouadizerwas{V. Barger, K. Cheung, A. Djouadi, B. Kniehl,
and P.M. Zerwas, \prdj{49} (1994) 79.}
(see also Ref.~[\djouadizerwas]) and it
is evident from Eq.~\couplings\ that
determination of the CP of a pure eigenstate requires transversely
polarized colliding photons and measurement
of the transverse polarization asymmetry
between event rates for parallel linear polarizations versus
perpendicular linear polarizations of the colliding photons.
As discussed in Refs.~[\gg,\gerice-\djouadizerwas], and detailed here,
the photons coming from the Compton back-scattering process can at
best be only partially polarized in the transverse direction. We will
discuss means for optimizing the observability of the transverse polarization
asymmetry by adjusting machine energy, electron and positron polarizations,
laser energy and laser polarizations.
\Ref\zerwasnew{As we were finalizing this paper, an overlapping work
by M. Kramer, J. Kuhn, M.L. Stong, and P.M. Zerwas,
appeared on the April 19, 1994 phenomenology bulletin board list:
hep-ph/9404280.}

For near optimal choices, we
compute the degree to which the CP-eigenvalues of each of the three
neutral Higgs bosons will be measurable; results are given as a function of the
fundamental parameters which determine the MSSM Higgs boson masses and
decays. For each of the Higgs bosons we survey the most important final state
channels for this determination. We find that the chances for measurement
of the CP-eigenvalue of the $\hl$ are generally good; prospects
in the case of the $\hh$ and $\ha$ are much more dependent upon model
parameters, but can also be good. Overall, this technique should prove very
valuable, especially if it is kept
in mind that CP determination will only be attempted after the
Higgs boson(s) have already been observed, \ie\ as part of a second generation,
hopefully high luminosity experiment.

\vskip .15in
\noindent{\bf 2. Procedure and Results}
\vskip .075in

The event rate for $\gam\gam\rta\h\rta X$ where $\h$ is the neutral Higgs boson
of interest and $X$ is a particular final state can be written
$$\eqalign{
{dN\over dt}(\gam\gam\rta\h\rta X)=&8\pi{BR(\h\rta X)\over E_{\epem}\mh^2}
\tan^{-1} {\Gamma_{res}\over \Gamma_{\h}}\Gamma(\h\rta \gam\gam)
{d\call_{\gam\gam}\over dy}\vert_{y_{\h}}\cr
\times&\left\{1+\vev{\xi_2\wt\xi_2}+
(\vev{\xi_3\wt\xi_3}-\vev{\xi_1\wt\xi_1})
\cala_3\right\}\cr \equiv &{d\over
dt}\left(N_{1+\vev{\xi_2\wt\xi_2}}+\cos(2\kappa)N_{\cala_3}\right)\, .\cr}
\eqn\dndt$$
In Eq.~\dndt, $BR(\h\rta X$) is the $\h\rta X$ branching ratio,
$\Gamma(\h\rta\gam \gam )$ is the $\h\rta\gam\gam$ width, $\Gamma_{\h}$ is the
total Higgs width,  $\mh$ is the Higgs mass, $E_{\epem}$ is the
center-of-momentum (CM) energy of  the $\epem$ collider, and we have given the
result appropriate for a  CP-conserving Higgs sector.  The asymmetry
$\cala_3$ is that defined in Ref.~[\gg], and will be given explicitly shortly.
For the $Q\anti Q$ channels, which have a
substantial background, we take $\Gamma_{res}$
to be either the best achievable experimental resolution, $\Gamma_{exp}$, or
$\Gamma_{\h}$, whichever is larger.
This minimizes the background level.  The other
channels considered are largely background free, and $\Gamma_{res}$ is
taken to be large (implying $\tan^{-1}\sim\pi/2$). In Eq.~\dndt,
${d\call_{\gam\gam}\over dy}\vert_{y_{\h}}$  is the $\gam\gam$-luminosity
averaged
over collisions for fixed $\epem$ CM energy, evaluated at $y_{\h}\equiv
E_{\gam\gam}/ E_{\ep\em}= \mh/E_{\ep\em}$;  $\kappa$ is the angle between the
directions of maximum linear  polarization of the two laser beams, assumed to
be
approaching nearly head-on; and $\xi_i$ and $\wt\xi_j$  ($i,j = 1,2,3$)
are the Stokes parameters which specify the polarization
state of the photons $\gam$ and $\wt\gam$,
respectively. For more details on the
notation see Ref.~[\telnovii]. The Stokes parameters
$\xi_{1,3}$ and $\wt \xi_{1,3}$
depend on $\kappa$; we have extracted the form of this dependence explicitly in
defining $N_{\cala_3}$. In our notation,
the brackets $\vev{...}$  denote averaging
over collisions at fixed $y_{\h}$
(which differs from the meaning of $\vev{...}$ in
Ref.~[\telnovi]).

The asymmetry $\cala_3$ can be written in terms of either linear polarization
or helicity amplitudes for Higgs production in $\gam\gam$
collisions: $$\cala_3\equiv
{\vert\calm_{\parallel}\vert^2-\vert\calm_{\perp}\vert^2
\over \vert\calm_{\parallel}\vert^2+\vert\calm_{\perp}\vert^2}=
{2{\rm Re\,}(\calm_{--}^*\calm_{++}) \over
\vert\calm_{++}\vert^2+\vert\calm_{--}\vert^2}\,,\eqn\aiiidef$$
where the $\calm$'s
are the amplitudes for photons of the indicated relative
linear polarizations or helicities. As already noted in association
with Eq.~\couplings,
$\cala_3$ is determined by the CP-eigenvalue of the Higgs boson produced;
$\cala_3 = +1(-1)$ for  CP-even
(-odd) scalars.\refmark\gg\
We wish to isolate $\cala_3$ in the event rate in order to determine the CP of
an observed $\h$.  This is made possible by the $\kappa$ dependence appearing
in
Eq.~\dndt\ --- the $\cala_3$ term depends upon $\cos(2\kappa)$, whereas
the $1 + \vev{\xi_2\wt\xi_2}$ term is independent of $\kappa$.
$\cala_3$ is thus isolated by taking the difference between event rates at
$\kappa = 0$ and $\kappa = {\pi\over 2}$, \ie\ parallel vs. perpendicular
linear polarizations for the initial laser beams.

Regarding the Higgs widths, we note that
for the MSSM Higgs bosons, $\Gamma_{\h}$
is typically much smaller than expected experimental resolutions,
although for large values of $\tanb$ the $b\anti b$ coupling
is enhanced and the associated decay width
can be significant for the $\ha$ and either the $\hl$ (if $\mha$ is small)
or the $\hh$ (if $\mha$ is large). Also, when decay into $t\anti t$
is kinematically allowed the $\ha$ and $\hh$ widths can be significant
even if $\tanb$ is not large.  Nonetheless, these widths are never much
larger than 5 to 10 GeV for the parameter ranges
considered here; our choice of $\Gamma_{exp}=15\gev$ is generally larger.
However, $\tan^{-1}[\Gamma_{exp}/\Gamma_{\h}]$ in Eq.~\dndt\
should not be approximated by simply using the $\pi/2$ limit.

There are two sources of background to be considered.  One is due to the
$1 + \vev{\xi_2\wt\xi_2}$ term for the Higgs itself which induces an error
in both of the accumulated event numbers, $N\vert_{\kappa = 0}$ and
$N\vert_{\kappa = {\pi\over 2}}$.
The second derives from continuum background(s) in the channel(s) in which
the $\h$ is detected (which are $\kappa$ independent
assuming a sum over all events subject only
to rapidity and/or polar angle cuts).  One measure of the
significance of the $\cala_3$ signal is
the number of standard deviations by which
$(N\vert_{\kappa = 0} - N\vert_{\kappa = {\pi\over 2}})$
exceeds the expected uncertainty of the combined background distribution:
$$\nsd (\cala_3)={{\sqrt{2} N_{\cala_3}} \over\sqrt{N_{1+\vev{\xi_2\wt\xi_2}}+
N_{cont}}}\, ,\eqn\nsdef$$
where $N_{\cala_3}$ and $N_{1+\vev{\xi_2\wt\xi_2}}$ are defined by
Eq.~\dndt, and
$N_{cont}$ is the event number for the continuum background.
We employ $\nsd (\cala_3)$ to determine the
range of MSSM Higgs sector parameters for which the
CP-eigenvalue of an MSSM $\h$
might be measurable.   We shall
evaluate $\nsd(\cala_3)$ for a yearly integrated $\gam\gam$ luminosity
of $20\fbi$ (a frequently employed canonical value for the first few years
of operation of a back-scattered laser beam facility).

A crucial issue is the best means for maximizing $\nsdaiii$.
$\naiii$ will be largest when the transverse polarizations of the colliding
photons are as big as possible.  For a given machine configuration
this means employing laser beams with maximal transverse
polarizations; we assume that essentially perfect transverse polarization
will be possible: $P_T(\gam_1)=P_T(\gam_2)=1$.
(Our notation is that $\gam_1$ collides with the $e^-$ and gives rise
to back-scattered photon $\gam$, while $\gam_2$ collides
with the $e^+$ producing $\wt\gam$. The associated Stokes parameters
are $\xi_i$ and $\wt\xi_i$, respectively.)
For $P_T(\gam_1)=1$ the value of $\xi_3$, for laser polarization orientation
such that $\xi_1=0$, is given by
$$\xi_3={2r^2\over (1-z)^{-1}+(1-z)-4r(1-r)}\,,\eqn\xithree$$
where $r=zx^{-1}/(1-z)$ with $z=E_{\gam}/E_{e^-}$.  A similar
result applies for $\wt\xi_3$. The parameter $x$ in the algebraic form
for $r$ is determined by
the $\epem$ CM energy, $E_{\ep\em}$, and the photon energy of the
laser, $\omega_0$:
$x={2E_{\ep\em}\omega_0\over m_e^2}$. We assume in the following that
both lasers are operated at the same $x$ value;
$x$ must be less than $4.83$ in order
that the back-scattering process be below the pair production threshold.
The maximum value of $z$ occurs for $r=1$, \ie\ $z_{max}=x/(1+x)$, at which
point $\xi_3$ reaches its largest value, $\xi_3(z_{max})=2(1+x)/[1+(1+x)^2]$.
This obviously suggests that choosing small values of $x$,
and operating at a machine energy that demands $z$ values near $z_{max}$
for a given (known) Higgs mass, will yield optimal results.
However, choosing a machine energy (assuming for the moment
that adequate machine energy is available) such that $y_{\h}=x/(1+x)$, which
would force $z=\wt z=x/(1+x)$, is too extreme since the folded luminosity
function would be zero. We return to this point below. First, we also note
that from the form of $\xi_3(z_{max})$ as a function of $x$ it is clear that
there is diminishing return in lowering $x$ much below 1.

 \TABLE\yoptimaltable{}

In this paper, we shall assume a minimum possible value for $x$
of $\xmin=0.5$, but that, when necessary, the lasers can be operated
at higher values of $x$ (up to the $x\sim 4.8$ pair threshold point).
Assuming for the moment no limitation on machine energy,
we have searched for the choice of
$y_{\h}$ that maximizes $\nsdaiii$ for a given value of $\xmin$.  The
results are most easily summarized in the form
$$\yhopt(\xmin)=\alpha(\xmin) {\xmin\over(1+\xmin)}\,.\eqn\yoptimal$$
The values of $\alpha(\xmin)$ as we vary $\xmin$ from 4.8 down to .2
in steps of .1
are given in Table~\yoptimaltable.  We see that the optimal
value of $y_{\h}$ is never much below the kinematic maximum.
In our results we shall always
employ a machine energy given by $\eepemopt=m_{\h}/\yhopt(\xmin)$
whenever this energy does not exceed the maximum machine energy
assumed available, denoted by $\eepemmach$.
For our particular choice of $\xmin=0.5$, $\yhopt(\xmin)\sim 0.299$.
Using this, the defining formula for $x$ with $x=\xmin=0.5$, and
$\eepem$ set equal
to $\eepemopt$, we find that the laser energy $\omega_0$ is given
in terms of the Higgs mass (in GeV) by:
$$\omega_0\sim {19.5~{\rm eV}\over \mh({\rm GeV})}\,,\eqn\omegalaser$$
\ie\ lasers with energies in the fractional eV range would generally
be necessary to operate at the optimal point.

 \midinsert
 \titlestyle{\tenpoint
 Table \yoptimaltable: We tabulate the values of $\alpha(x)$, Eq.~\yoptimal,
for $x$ ranging from 4.8 down to .2 in steps of .1.
}
 \smallskip

 \thicksize=0pt
 \hrule \vskip .04in \hrule
 \begintable
0.966 & 0.966 & 0.966 & 0.964 & 0.964 & 0.964 & 0.962 & 0.962 & 0.962 & 0.960
& 0.960 & 0.960 \nr
0.958 & 0.958 & 0.958 & 0.956 & 0.956 & 0.954 & 0.954 & 0.952 & 0.952 & 0.950
& 0.949 & 0.948 \nr
0.947 & 0.946 & 0.944 & 0.942 & 0.940 & 0.938 & 0.937 & 0.936 & 0.934 & 0.932
& 0.928 & 0.926 \nr
0.924 & 0.920 & 0.918 & 0.914 & 0.910 & 0.906 & 0.902 & 0.898 & 0.892 & 0.888
& 0.884 & $-$
\endtable
 \hrule \vskip .04in \hrule
 \endinsert

For larger $\mh$ values, it is frequently the case that
$\eepemopt>\eepemmach$.  In this situation,
we have explicitly searched for the compromise
values of $x$ and $y_{\h}$ that maximize $\nsdaiii$, subject to
the restrictions $\eepem\leq\eepemmach$, $\xmin\leq x\leq 4.8$.
For all such cases considered
we have found that it is always best to employ $\eepem=\eepemmach$.
The best choice of $x$ is situation dependent, but does tend
to be fairly near the minimum possible choice of $x=\yhmach/(1-\yhmach)$,
where $\yhmach=\mh/\eepemmach$.

Two additional notes are useful.  First,
the optimal choices outlined above are independent of the particular
Higgs boson decay channel being considered. Second,
the best choice for $y_{\h}$ is always fairly close to its kinematical
limit.  This means that the $\gam\gam$ subprocess energy is
such that $Q\anti Q$ backgrounds arising from ``resolved-photons''
\Ref\resolved{See, for example, O.J.P. Eboli, M.C. Gonzales-Garcia,
F. Halzen, and D. Zeppenfeld, \prdj{48} (1993) 1430.}\
will be very suppressed compared to the direct $\gam\gam\rta
Q\anti Q$ subprocess continuum backgrounds that we shall consider.
This is simply due to the fact that if an incoming photon radiates
a secondary particle which participates in $Q\anti Q$ production,
there is a very low probability that the $Q\anti Q$ pair
will have invariant mass large enough to be confused with a
signal at the (known) Higgs mass.

Finally, a small amount of optimization of $\nsdaiii$ is possible
by choosing $e^-$ and $e^+$ helicities such that
$1+\vev{\xi_2\wt\xi_2}$ is as small as
possible. (This turns out to be appropriate even in the $b\anti b$
and $t\anti t$ channels where the continuum backgrounds increase
with decreasing $1+\vev{\xi_2\wt\xi_2}$.)  Our results are given
for $\vev{\lam(e^-)}=-\vev{\lam(e^+)}=0.45$. However the
quoted $\nsdaiii$ values are
decreased by at most 10\% if the $e^-$ and $e^+$ beams are unpolarized.

We now turn to quantitative results.  In the case of the $Q\anti Q$ channels
we employ an experimental mass resolution in
the final state of $\Gamma_{exp}=15\gev$. We give results assuming the rumored
CDF top quark mass value of 175 GeV.
For the $Q\anti Q$ final states, we impose an angular
cut in the $\gam\gam$ CM frame of $\vert\cos\theta\vert<0.8$. We then
compute $\nsdaiii$ as a function of the MSSM parameter $\mha$, varying $\mha$
between 40 and 800 GeV, for the two representative values of $\tanb = 2,20$.
Radiative corrections for the $\hl$ and $\hh$ masses are computed
neglecting squark mixing.  Initially, we shall employ $\mstop=1\tev$.
Alterations in our results for lower values of the squark masses will
be summarized towards the end of the paper.

Below, we summarize results for the decay channels with highest $\nsdaiii$,
using an integrated photon-photon luminosity of $L=20\fbi$.
While such an integrated
luminosity will not always turn out to be adequate for the determination
of $\cala_3$, it is to
be hoped that higher luminosities could eventually be accumulated.  Since
both signal and background scale as $L$, $\nsdaiii$ scales as $\sqrt L$.

Before beginning our survey, it is useful to recall
some basic features of the decays of the Higgs bosons
(see Refs.~[\hhg] and [\gerice] for more detailed discussions) in order
to better understand why particular channels yield the best $\nsdaiii$ values.
First, we emphasize that we have considered in this work only those
channels that contain standard model particles.
Once more is known experimentally about the supersymmetric spectrum (for
which an $\epem$ collider should be an excellent machine)
the SUSY decay modes for
the Higgs bosons can be computed, and a reassessment of all channels, including
SUSY channels, performed. We shall discuss the modifications
in SM channels for a sample case in which SUSY modes are allowed.
A study of the possibilities for measuring $\cala_3$ using SUSY
decay modes is beyond the scope of this study. Possible problems include
invisible decay modes such as $\h\rta \wt\chi_1^0 \wt\chi_1^0$, and, more
generally, the presence of missing energy that could worsen the experimental
resolution in a typical SUSY decay channel, making it more difficult to
reject background events.
For our basic scenario we shall adopt a universal soft SUSY-breaking squark
mass of $1\tev$, and the masses of charginos and neutralinos will
be set to very large values by the soft-SUSY breaking
parameter choices $M=-\mu=1\tev$. Squark mixing will be neglected.

In the absence of SUSY modes,  decays
of the $\hl$ are dominated by the
$b\anti b$ channel, independent of the value of
$\tanb$. ($\mhl$ is always smaller than $2\mt$.) In contrast,
the decays of the $\ha$ and $\hh$ change
substantially as $\tanb$ is increased.  For
small to moderate $\tanb$, the primary
$\ha$ decays are to $b\anti b$ so long as
$\mha< \mz+\mhl$. However, once the $\ha$
mass is above the $Z\hl$ threshold, $\ha\rta Z\hl$
decays tend to dominate, although
the $b\anti b$ channel still remains significant. For small to moderate
$\tanb$,
the $\hh$ decays primarily to $b\anti b$ for $\mhh<2\mhl$,
while $\hh\rta \hl\hl$
decays are dominant once $\mhh>2\mhl$. Of course, for $\mha,\mhh$ above $2\mt$,
$\ha,\hh\rta t\anti t$ decays dominate for small to moderate $\tanb$.
At large $\tanb$, the couplings of the
$\ha$ and $\hh$ to $b\anti b$ and $\tau^+\tau^-$
become highly enhanced, and these
two modes dominate with the $b\anti b$ channel
having about 90\% branching ratio in the absence of SUSY decay channels.

\FIG\hlfig{
$\nsdaiii$ for the $\hl$ is displayed as a function of $\mhl$
for $\tanb=2$ and 20 for the $b\anti b$ final state, assuming $L=20\fbi$.
The solid (dotted) curves are for $\xmin=0.5(4.0)$ with $\eepem=\eepemopt$.
}
\midinsert
\vbox{\phantom{0}\vskip 5.0in
\phantom{0}
\vskip .5in
\hskip -0pt
\special{ insert scr:a3hl.ps}
\vskip -1.45in }
\centerline{\vbox{\hsize=12.4cm
\Tenpoint
\baselineskip=12pt
\noindent Figure~\hlfig:
$\nsdaiii$ for the $\hl$ is displayed as a function of $\mhl$
for $\tanb=2$ and 20 for the $b\anti b$ final state, assuming $L=20\fbi$.
The solid (dotted) curves are for $\xmin=0.5(4.0)$ with $\eepem=\eepemopt$.
}}
\endinsert

Non-Higgs backgrounds to these modes are negligible
except for the $b\anti b$ and $t\anti t$ channels, for which
we include the continuum $\gam\gam\rta b\anti b$ and $\gam\gam\rta t\anti t$
processes, respectively. As discussed earlier,
resolved photon backgrounds are insignificant since the optimal
$\gam\gam$ collision energies for measurement of $\cala_3$ are always
close to the kinematical limit.

Our results appear in a series of figures.  Let us focus first on the $\hl$.
For $\mt=175\gev$ the $\hl$ has mass below $100(122.5)\gev$
for $\tanb=2(20)$ when $\mstop=1\tev$.  Thus, for $\yhopt(\xmin=0.5)=0.299$
the optimal machine energy always lies below about $410\gev$.
Assuming an available (but tunable) energy of $\eepemmach=500\gev$, this means
that we can always operate at the optimal point.
Only the $b\anti b$ final state is of relevance, for which the values
of $\nsdaiii$ appear as the solid lines in Fig.~\hlfig.  Had
we adopted the much less optimal choice of $\xmin=4.0$ the resulting
$\nsdaiii$ values would be those indicated by the dotted lines in Fig.~\hlfig,
\ie\ about a factor of 5 worse.  In contrast, results for $\xmin=1$ are
only slightly worse than for $\xmin=0.5$.  From this we conclude
that if $\xmin$ values of order 0.5 to 1 are possible,
there is quite a resonable chance of being able to determine $\cala_3$
in the case of the $\hl$. Only low $\mhl$ values at low to moderate
$\tanb$ values would appear to require integrated luminosities above
$L\sim 20\fbi$. In contrast, if lasers of
adequate power and intensity with energies in the fractional eV
range are not possible, much larger integrated luminosity would be required
over a broad range of parameter space; $\xmin=4.0$ would
require $L\gsim 200\fbi$ in order to measure $\cala_3$ for most $\mhl$
values.

\FIG\hafiglow{
$\nsdaiii$ for the $\ha$ is displayed as a function of $\mha$
for $\tanb=2$ and 20 for the $b\anti b$ (solid), $Z\hl$ (dots)
and $t\anti t$ (dashes) final states, assuming $L=20\fbi$,
a maximum machine energy of $\eepemmach=500\gev$, and $\xmin=0.5$.
}
\midinsert
\vbox{\phantom{0}\vskip 5.0in
\phantom{0}
\vskip .5in
\hskip -0pt
\special{ insert scr:a3halow.ps}
\vskip -1.45in }
\centerline{\vbox{\hsize=12.4cm
\Tenpoint
\baselineskip=12pt
\noindent Figure~\hafiglow:
$\nsdaiii$ for the $\ha$ is displayed as a function of $\mha$
for $\tanb=2$ and 20 for the $b\anti b$ (solid), $Z\hl$ (dots)
and $t\anti t$ (dashes) final states, assuming $L=20\fbi$,
a maximum machine energy of $\eepemmach=500\gev$, and $\xmin=0.5$.
}}
\endinsert

\FIG\hafighigh{
$\nsdaiii$ for the $\ha$ is displayed as a function of $\mha$
for $\tanb=2$ and 20 for the $b\anti b$ (solid), $Z\hl$ (dots)
and $t\anti t$ (dashes) final states, assuming $L=20\fbi$,
a maximum machine energy of $\eepemmach=1.5\tev$, and $\xmin=0.5$.
}
\midinsert
\vbox{\phantom{0}\vskip 5.0in
\phantom{0}
\vskip .5in
\hskip -0pt
\special{ insert scr:a3hahigh.ps}
\vskip -1.45in }
\centerline{\vbox{\hsize=12.4cm
\Tenpoint
\baselineskip=12pt
\noindent Figure~\hafighigh:
$\nsdaiii$ for the $\ha$ is displayed as a function of $\mha$
for $\tanb=2$ and 20 for the $b\anti b$ (solid), $Z\hl$ (dots)
and $t\anti t$ (dashes) final states, assuming $L=20\fbi$,
a maximum machine energy of $\eepemmach=1.5\tev$, and $\xmin=0.5$.
}}
\endinsert

Turning now to the $\ha$, we present results for $\mha\geq40\gev$.
Exactly how high in $\mha$ we can go and the range of $\mha$ for
which we can employ $x=\xmin=0.5$ depends upon the available
machine energy, $\eepemmach$. We contrast the two cases
of $\eepemmach=500\gev$ and $\eepemmach=1.5\tev$ in Figs.~\hafiglow\
and \hafighigh, respectively. At $\eepemmach=500\gev$, one can
adjust $\eepem$ so as to employ $\yhopt(\xmin=0.5)$
so long as $\mha\lsim 150\gev$,
but beyond that $x$ must be increased above $\xmin=0.5$; we have
searched for the optimal choice as described earlier. Kinematically,
the maximum $\mha$ that can be probed at $\eepemmach=500\gev$
is just above $400\gev$ if $x$ lies below 4.83
(\ie\ below the pair-production threshold).
{}From Fig.~\hafiglow\ we see that at low
$\tanb$ the ability to measure $\cala_3$ declines rapidly as soon as
we pass the $t\anti t$ threshold at $\mha\sim 350\gev$, but that below
that $\nsdaiii$ is significant in the $Z\hl$ channel.  (In drawing
this conclusion, we have assumed that this channel is relatively background
free in all possible final state modes.)  Determination of $\cala_3$
in the $b\anti b$ channel at low $\tanb$ would require enhanced luminosity.
At high $\tanb$, Fig.~\hafiglow\ shows that the $\mha$ range over
which $\cala_3$ can be measured is considerably diminished.  Only
the $b\anti b$ final state is useful, and only for $\mha\lsim 70\gev$.
Increased luminosity leads to only a moderate increase in the maximum
$\mha$ for which $\cala_3$ could be measured; for instance,
an accumulated luminosity of $L\sim 200\fbi$ would extend the $\nsdaiii>3$
region only to $\mha\sim 140\gev$.

At $\eepemmach=1.5\tev$, increasing $x$ above $\xmin=0.5$ only becomes
necessary
for $\mha\gsim 450\gev$. Thus, the $Z\hl$ mode allows even larger
$\nsdaiii$ values for $\mha$ near $2\mt$ at low $\tanb$ (see Fig.~\hafighigh),
and even for $\mha$ somewhat above $2\mt$ determination of $\cala_3$ in
the $t\anti t$ mode will be possible. The exact reach will depend upon
the available integrated luminosity, extending
for $L=200\fbi$ to possibly as high as $\mha\sim 700\gev$.
Unfortunately, prospects for $\cala_3$
determination at high $\tanb$ remain quite limited, despite the
increased machine energy. Even for $L=200\fbi$ the $\nsdaiii>3$ region
is confined to $\mha\lsim 160\gev$.
It should be noted that increasing $\eepemmach$
still further does not significantly increase the maximum $\mha$ values
for which $\cala_3$ can be measured for either low or high $\tanb$.

\FIG\hhfighigh{
$\nsdaiii$ for the $\hh$ is displayed as a function of $\mhh$
for $\tanb=2$ and 20 for the $b\anti b$ (solid), $ZZ$ (non-$4\nu$) (dots),
$\hl\hl$ (dotdash) and $t\anti t$ (dashes) final states, assuming $L=20\fbi$,
a maximum machine energy of $\eepemmach=1.5\tev$, and $\xmin=0.5$.
}
\midinsert
\vbox{\phantom{0}\vskip 5.0in
\phantom{0}
\vskip .5in
\hskip -0pt
\special{ insert scr:a3hhhigh.ps}
\vskip -1.45in }
\centerline{\vbox{\hsize=12.4cm
\Tenpoint
\baselineskip=12pt
\noindent Figure~\hhfighigh:
$\nsdaiii$ for the $\hh$ is displayed as a function of $\mhh$
for $\tanb=2$ and 20 for the $b\anti b$ (solid), $ZZ$ (non-$4\nu$) (dots),
$\hl\hl$ (dotdash) and $t\anti t$ (dashes) final states, assuming $L=20\fbi$,
a maximum machine energy of $\eepemmach=1.5\tev$, and $\xmin=0.5$.
}}
\endinsert

Finally, let us consider the $\hh$. Results for $\nsdaiii$ at
$\eepemmach=1.5\tev$ and $\xmin=0.5$ are displayed in Fig.~\hhfighigh,
assuming $L=20\fbi$. We see that $\cala_3$ will be
most observable in the $\hl\hl$ channel when $\tanb$ is not large
and $\mhh\lsim 2\mt$.  When $\mhh$ is very near its lower bound,
the $\ha\ha$ channel also becomes useful (although it is not displayed).
For $\mhh\gsim 2\mt$ the $t\anti t$ channel may provide the best
opportunity for low $\tanb$, but $L$ above $20\fbi$ will
generally be required even for $\xmin=0.5$.
The values of $\nsdaiii$ for the $ZZ$
channel are given after multiplying rates by $1-[BR(Z\rta\nu\anti\nu)]^2$
(\ie\ so as to avoid the totally invisible final state mode), but do
not include the significant one-loop $ZZ$ continuum background.
\Ref\jikia{G.V. Jikia, \plbj{298} (1993) 224.}\
Inclusion of this continuum background would further diminish the prospects
for probing $\cala_3$ in the $ZZ$ channel.
At high $\tanb$, determination of $\cala_3$
for the $\hh$ will be very difficult unless the $\hh$ is very near
its lower mass limit (corresponding to small $\mha$), in which case
one of the three channels --- $b\anti b$, $\hl\hl$ or $\ha\ha$ ---
provides an $\nsdaiii$ value in the range between 5 and 25.
(We have not attempted to display this.)

Although not presented, the $\eepemmach=500\gev$ figure for the $\hh$
exhibits differences from Fig.~\hhfighigh\ closely analogous to
those between Fig.~\hafiglow\ and \hafighigh\ for the $\ha$.
The most important difference in the case of the
$\hh$ is that for $\tanb=2$ the $\nsdaiii$ value for the $\hl\hl$
final state mode declines from the value of $\sim 5$
shown in Fig.~\hhfighigh\ to $\sim 2.5$ just below the $t\anti t$ threshold.

This concludes our discussion of the basic scenario in which all
SUSY partner particles, in particular squarks and inos, are taken
to be very heavy, \ie\ in the 1 TeV mass range.
Three variants on this basic scenario have been considered.
In the first variant ($V1$), we have decreased the universal soft squark
mass parameter to 350 GeV, maintaining large ino masses by continuing to take
$M=-\mu=1\tev$.  In the second case ($V2$), we have kept the
soft squark mass large ($1\tev$), while allowing small ino masses
as determined by $M=-\mu=150\gev$.  In the third variant ($V3$)
the soft squark mass is taken to be 350 GeV {\it and} we have set
$M=-\mu=150\gev$. The first case, $V1$, is interesting
in that the radiative corrections to the Higgs masses are greatly reduced,
while in $V2$ the inos are light enough to reduce the SM
channel branching fractions for heavier Higgs bosons, while the charginos
are light enough to significantly contribute to the one-loop $\gam\gam$
couplings of the different Higgs bosons. Variant $V3$ combines both
effects. We summarize below the effect of the variants $V1$ and $V2$
on the achievable $\nsdaiii$ values, assuming a machine energy of
$\eepemmach=1.5\tev$. Results for variant $V3$ are very similar to those
for $V2$ and will not be discussed in detail.

\medskip
\nou{$\hl$}
\medskip

\nou{$V1$} The maximum $\mhl$ value declines, but the achievable
$\nsdaiii$ values in the $b\anti b$ channel are little affected
(at a given $\mhl$ value). This is true at both low and high $\tanb$.

\nou{$V2$} The achievable $\nsdaiii$ values decline somewhat (due
to the chargino loops cancelling some of the $W$-loop contribution
to the $\gam\gam$ coupling of the $\hl$), but never by more than about 30\%.

\medskip
\nou{$\ha$}
\medskip

\nou{$V1$} Decreasing the soft squark mass has very little impact
upon $\nsdaiii$ at either small or large $\tanb$.  The largest effect is
a $\sim 20\%$ decrease in $\nsdaiii$ in the $b\anti b$ and $Z\hl$
channels in the $200\gev \lsim \mha\lsim 2\mt$ range at low $\tanb$.
This is primarily due to small squark-loop contributions to the
$\gam\gam$ coupling of the $\ha$.

\nou{$V2$} This variant has a major impact at both small and large $\tanb$.
At $\tanb=2$, $\ha$ decays to ino pairs become significant and
one-loop contributions to the $\gam\gam$ coupling of the $\ha$ are
also important.  The net effect is to decrease $\nsdaiii$ in all
channels, with the maximum result in the $Z\hl$ ($t\anti t$)
channel at $\mha\lsim 2\mt(\gsim 2\mt)$
being reduced to $\nsdaiii\sim 2(3.5)$, at $L=20\fbi$.
Meanwhile, $\nsdaiii$ in the $b\anti b$ channel never rises above 0.4
at $\tanb=2$. At $\tanb=20$, the coupling of the $\ha$
to $b\anti b$ is enhanced, and $BR(\ha\rta b\anti b)$ remains
large despite the presence of ino-pair channels.  Meanwhile, the chargino
loops yield a significant increase in the $\gam\gam$ coupling
of the $\ha$.  The result is that $\nsdaiii$ remains above 2
in the $b\anti b$ channel for $\mha$ up to $\mha\sim 400\gev$,
with a large peak at $\mha\sim 220$ where $\nsdaiii\sim 8$.

\medskip
\nou{$\hh$}
\medskip

\nou{$V1$} As for the $\ha$, decreasing the squark mass to 350 GeV
has relatively small impact upon $\nsdaiii$.  At $\tanb=2$, the $\nsdaiii$
values achievable in the $\hl\hl$ channel for $\mhh<2\mt$ decline
slightly (by at most 20\%), while $\nsdaiii$ increases by up to 50\%
in the $t\anti t$ channel for $\mha$ within 100 GeV of $2\mt$.
At $\tanb=20$,  only a tiny decrease in $\nsdaiii$ for the $b\anti b$
channel results from decreasing the squark mass.

\nou{$V2$} The impact of lowering the $M=-\mu$ value to 150 GeV
is once again quite significant.  At $\tanb=2$, $\nsdaiii$ in the $\hl\hl$
channel falls below 1 for $\mhh\gsim 500\gev$ (as opposed to $\gsim 600\gev$
for our standard scenario). In the $t\anti t$ channel, $\nsdaiii$
only barely reaches 0.9 at its maximum point (as compared to $\sim 2.2$
in the standard scenario).  In contrast, at $\tanb=20$, $\nsdaiii$
is enhanced in the $b\anti b$ channel (just as in the $\ha$ case),
remaining above 1 out to $\mhh\sim 500\gev$ with a broad peak
in the vicinity of $\mhh\sim 300\gev$ with maximum of $\nsdaiii\sim 2.2$.

\medskip
\nou{A Fourth Generation}
\medskip

As a final variant, one might consider the addition of a fourth
generation.  This can have a dramatic effect. At large $\tanb$,
the observability of $\cala_3$ is dramatically increased for all
three neutral Higgs bosons. At small $\tanb$ the impact is more varied.
Observability of the $\hl$ is decreased (due to cancellation of fourth
generation loops against the $W$-loop contribution to the $\gam\gam$ coupling
of the $\hl$).  In contrast, observability of the $\ha$ at small $\tanb$
is dramatically increased since the fourth generation loop contributions to the
$\gam\gam$ coupling of the $\ha$ add to the top-quark
loop (and the $W$ loop is absent).
Meanwhile, at small $\tanb$, $\nsdaiii$ for the $\hh$ is not dramatically
altered by the addition of a fourth generation.

\vskip .15in
\noindent{\bf 3. Conclusion}
\vskip .075in

The outlook for measuring the CP-eigenvalues of the three neutral MSSM Higgs
bosons via back-scattered photons depends tremendously on the ability
to build lasers with high luminosity at low photon energies.  At low
Higgs mass values, it will be important to be able to adjust the
laser energy ($\omega_0$) and electron/positron
beam energies so as to be near the optimum values for $y_{\h}\equiv\mh/\eepem$
while retaining $x\equiv {2\eepem\omega_0\over m_e^2}\lsim 1$
in order to have large transverse polarization
for the back-scattered photons. For larger values of
$\yhmach\equiv\mh/\eepemmach$ (where
$\eepemmach$ is the maximum available $\epem$ machine energy),
the ability to adjust the laser photon energy to yield $x$ values
not too much larger than the minimum required, $\yhmach/(1-\yhmach)$,
will again be vital.  Of course, we will presumably know in advance
the masses of the Higgs bosons, so that in practice only a discrete
set of laser energies would have to be available.

Provided fairly optimal choices for the laser and beam energies are
possible, prospects for measuring the vital asymmetry $\cala_3$ (which is $+1$
for a CP-even Higgs boson and $-1$ for a CP-odd Higgs) are good
in the case of the $\hl$ and are generally good for the $\ha$ and $\hh$
at low $\tanb$, even if only $20\fbi$ of integrated luminosity is accumulated.
At large $\tanb$, determination of $\cala_3$ for the $\ha$ and $\hh$
will generally require significantly
larger integrated luminosities (increasing for increasing mass)
for all but the lowest mass values.
These results (which assume heavy squarks, charginos and
neutralinos) are not much altered if squarks are
light. If inos are taken to be light some decline in the statistical
significance for a measurement of $\cala_3$ does occur in the
case of the $\ha$ and $\hh$ at small $\tanb$, while at large
$\tanb$ it generally becomes easier to determine $\cala_3$ for
the $\ha$ and $\hh$.
Finally, we have noted that the addition of a fourth generation
greatly boosts the ease with which $\cala_3$ can be determined
for the $\ha$ and $\hh$ (but can worsen the prospects for the $\hl$
if $\tanb$ is small).

While it is likely that the first indication of
whether a given neutral Higgs is CP-even
or CP-odd will come simply from the
size of its $Z^0\rta Z^0\h$ production cross
section (only a CP-even Higgs has a tree-level coupling of the required type),
use of $\cala_3$ appears to provide superior opportunities
in comparison to other options for {\it directly} determining the CP of
Higgs bosons.\Ref\othermeans{For a review of and references to other methods,
see Ref.~[\ghawaii] and M.L. Stong and K. Hagiwara,
Proceedings of the ``Workshop on Physics
and Experiments with Linear $\epem$ Colliders'', Waikaloa, Hawaii,
April 26-30 (1993), eds. F.A. Harris, S.L. Olsen, S. Pakvasa, and X. Tata,
p. 631; an updated review of some of these other methods
also appears in Ref.~[\zerwasnew].}

Finally, very high instantaneous photon-photon
luminosities appear to be technically
feasible, \Ref\highl{D. Borden, private communication.}
by running at correspondingly high instantaneous $\epem$ luminosities (higher
than can be employed without beam disruption \etc\
in direct $\epem$ collision studies). Although this mode of operation
would be expensive using current technology, a combination of money and/or
new technology could allow for photon-photon integrated luminosity
in the $\gsim 100\fbi$ domain. Measurement of $\cala_3$
would then be even more superior in comparison to other methods
for direct determination of the Higgs boson CP-eigenvalues, as referenced
above, that rely on the more limited integrated luminosity
that would be possible for direct $\epem$ collision studies.

\smallskip\centerline{\bf Acknowledgements}
\smallskip
This work has been supported in part by Department of Energy
grant \#DE-FG03-91ER40674
and by Texas National Research Laboratory grant \#RGFY93-330.
We are grateful to D. Borden and H. Haber for helpful discussions.

\smallskip
\refout
\end